# FUTURE WISHES AND CONSTRAINTS FROM THE EXPERIMENTS AT THE LHC FOR THE PROTON–PROTON PROGRAMME

R. Jacobsson, CERN, Geneva, Switzerland

*Abstract*

Hosting six different experiments at four different interaction points and widely different requirements for the running conditions, the LHC machine has been faced with a long list of challenges in the first three years of luminosity production (2010–2012, Run 1), many of which were potentially capable of limiting the performance due to instabilities resulting from the extremely high bunch brightness. Nonetheless, LHC met the challenges and performed extremely well at high efficiency and routinely with beam brightness at twice the design, well over one-third of the time in collision for physics, average luminosity lifetimes in excess of 10 h and extremely good background conditions in the experiments.

While the experimental running configurations remain largely the same for the future high luminosity proton–proton operational mode, the energy and the luminosity should increase significantly, making a prior assessment of related beam–beam effects extremely important to guarantee high performance. Of particular interest is the need for levelling the luminosity individually in the different experiments. Luminosity control as the more general version of 'levelling' has been at the heart of the success for LHCb, and to a large extent also for ALICE, throughout Run 1. With the increasing energy and potential luminosity, luminosity control may be required by all the experiments at some point in the future as a means of controlling the pileup conditions and trigger rates, but possibly also as a way of optimizing the integrated luminosity.

This paper reviews the various motivations and possibilities for controlling the luminosity from the experiments' point of view, and outlines the future running conditions and desiderata for the experiments as they are viewed currently, with the aim of giving guidelines for different options.

## INTRODUCTION

The LHC is special in that it has to cater for both direct discovery and precision physics. This means that the outcome of the LHC Run 1 should also be viewed against the backdrop of the rapidly evolving requirements from the different experiments, as the performance parameter space of the machine itself unfolded:

- Up to six experiments taking data simultaneously at four interaction points with instantaneous luminosity ranging from $2 \times 10^{30}$ cm$^{-2}$ s$^{-1}$ to $8 \times 10^{33}$ cm$^{-2}$ s$^{-1}$.
- Luminosity levelling at an intermediate luminosity of $4 \times 10^{32}$ cm$^{-2}$ s$^{-1}$ in one IP and levelling at a low luminosity of $2 \times 10^{30}$ cm$^{-2}$ s$^{-1}$ in another IP.
- 50 ns collision scheme with different number of colliding bunches in the different IPs.
- One IP shifted by 11.25 m with respect to the nominal IP.
- Non-colliding bunches and bunches colliding at an offset in only one IP.
- Collision scheme in one IP with main bunches against enhanced satellites from the LHC injectors in the 50 ns gaps.
- Two IPs with experimental spectrometer magnets and need for regular polarity changes.
- A 20° tilted crossing scheme in one IP to assure the same effective crossing angle in both polarities.

All these requirements effectively meant that the preparation for collisions in each fill was a very delicate process and occasional periods with instabilities from collective effects were not easy to deal with. On the other hand, these requirements together with the exceptional performance of the machine allowed exploration of the LHC parameter space for the future. There is a wealth of information available that is still under analyses to help decide on the best options for Run 2 and Run 3 (2015–2017 and 2019–2021, respectively, as the schedule currently stands).

Moreover, the fact that Run 1 already exposed the experiments to very high event pileup has also been very constructive in pushing the experiments to improve on the performance of the current detectors, trigger and reconstruction. It also allowed weaknesses to be revealed and choices to be guided for the successive upgrade programmes for all the experiments. The upgrade programmes are now well underway.

In terms of physics, the Run 1 high luminosity proton–proton programme left us with, on the one hand, the fundamental discovery of the existence of scalars in nature and a compatibility with a 126 GeV/$c^2$ Standard Model Higgs boson. On the other hand, the absence of a non-Standard Model signal in the precision measurements, in particular in the heavy flavour sector, and in the other direct searches for new particles, strongly indicate that New Physics is either very heavy or is only very weakly coupling to the Standard Model particles.

Of course, all of this does not go without saying that the proton–proton programme also provided vital physics input to the ion physics programme.

While the results above were clearly one of the very likely outcomes of Run 1, it gives a particular significance to the initial scope of Run 2 but leaves the scope beyond Run 2 largely open. That is to say, while the answers found in Run 1 hint at no New Physics but rather an even more well-established Standard Model as low-energy effective theory, the unsolved fundamental questions, such as the neutrino oscillations, the baryon asymmetry, the dark matter, dark energy, etc., are still as much in need of New Physics. As it stands now, precision measurements on the newly found boson are likely to be one of the best compasses to suggest the direction on New Physics.

Firstly, the nature of the newly found boson should be determined through precise measurements of the couplings to the vector bosons and the fermions, and its role in the electroweak symmetry breaking and mass generation need to be understood. Of particular importance is the question whether the newly found boson solves the unitarity violations in scattering amplitudes with triple and quartic W and Z boson couplings. This programme of work requires access to all the production and decay modes of the 126 GeV/$c^2$ boson, some of which are very challenging in an environment with high event pileup.

The other path in the search for New Physics will consist of continued precision measurements on rare heavy flavour decays and CP violation, and direct searches for new particles. The increase of the centre-of-mass energy in LHC to 13 TeV will almost double the heavy flavour production cross-sections with respect to 7–8 TeV, and the parton luminosities for the production of new particles at a mass scale of 1 TeV through gluon–gluon, Σquark–anti-quark, Σgluon–quark(anti-quark) interactions will increase by an order of magnitude and at 2 TeV by two orders of magnitude.

The current physics situation with the high priority on precision measurements means that LHC will be even more challenged to produce the largest possible integrated luminosity at the smallest possible event pileup.

With this in mind, this paper focuses on the options from the experiments' point of view for the future high luminosity proton–proton programme, which currently appears to present the machine with the most involved and the most challenging aspects with respect to beam–beam effects.

Table 1 shows the beam parameters assumed in this paper for discussing the preferred option of 25 ns operation. They stem from the very successful Bunch Compression and Merging Scheme (BCMS) [1,2] that was devised by the LHC injectors in the second half of 2012 to significantly improve the performance with the 25 ns beam. Note that the plots in this paper serve as illustrations. They do not show the ultimate performance. The exact future running conditions of the experiments is currently under study.

Table 1: Assumed beam parameters in the discussion of the preferred option for 25 ns operation [1, 2].

|       | $I_b$ | $\varepsilon_N$ [μm] | $N_b$ (IP1,2,5/IP8) | $\sqrt{s}$ [TeV] | $\sigma_z$ [cm] |
|-------|-------|----------------------|---------------------|------------------|-----------------|
| Run 2 | $1.15\times10^{11}$ | 1.9 | 2500 / 2400 | 13 | 10 |
| Run 3 | $1.4\times10^{11}$ | 1.9 | 2800 / 2600 | 13 | 10 |
| HL-LHC | $2.2\times10^{11}$ | 2.5 | 2800 / 2600 | 14 | 7.5 |

## MOTIVATIONS FOR LUMINOSITY AND PILEUP CONTROL

In the considerations of the strategy to maximize the LHC physics output, obviously neither the peak luminosity nor the delivered integrated luminosity is a good figure-of-merit. Several things may affect the data taking efficiency and the data quality. However, while the experiments use the full good-quality recorded integrated luminosity for physics analyses, not even this is a good figure-of-merit of the physics performance. The physics selection efficiencies and the background rejection, i.e. the physics yields, rely on observables whose resolutions are affected by many things, including in-time event pileup, out-of-time event pileup (also called 'spill-over'), background conditions, etc. The out-of-time pileup is due to the integration time of the detector components and typically implies that a certain fraction of the signal from the previous events is sampled with every event. The yield varies strongly between physics channels, depending on the type of final states and the angular distribution. For instance, final states with muons in the barrel are relatively easy to reconstruct even at very large event pileup, while channels with the need for reconstructing a whole decay chain of hadrons, including π/K identification in the forward region, are much more difficult (Fig. 1). Thus, whichever mode of operation is chosen, the proper figure-of-merit to maximize the physics output is the effective usable integrated luminosity. It will obviously involve a difficult and complicated combination of experimental priorities and compromises, which in the end results in the choice of a running condition and trigger configuration.

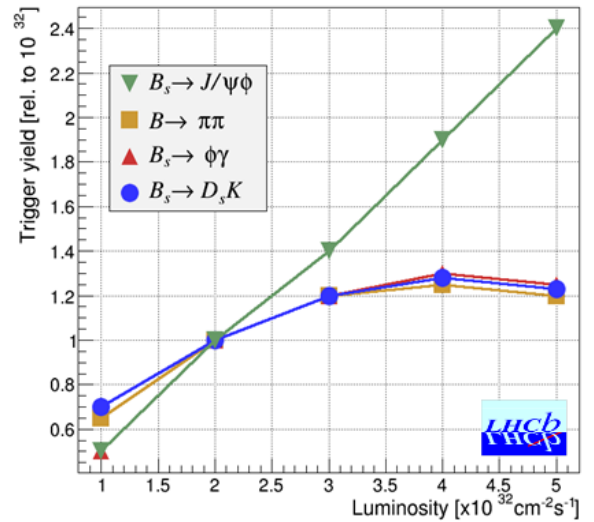

Figure 1: The plot shows the physics yield as a function of luminosity with simple final states such as muons, and the more complicated hadron final states which typically require reconstructing and the decay chain.

The following provides a list of the constraints and the effects which can potentially require limiting or levelling the instantaneous luminosity:

- **Instrumental limitation or instantaneous damage:** An instantaneous luminosity well beyond the design limits of some type of detectors may lead to such important ionization or changes of the dielectric properties that damage follows. For instance, gaseous detectors may receive damage from sparking. In most cases, the detectors will be well protected by current trip thresholds but it obviously involves relatively long recovery times which in turn affect the data taking efficiency and the data quality.
- **Detector ageing** is normally assumed to be linear in instantaneous luminosity and in integrated luminosity but the limits to this assumption is largely an unknown until it has been measured. This means that the detector replacement and upgrade programmes in place are based on experience and expected performance. It also means that the luminosity to which the detector is exposed should ideally be as close as possible to the effective usable luminosity.
- **Detector conditioning:** In some rare cases, detectors have shown a reduced performance or other negative effects at the immediate onset of a high instantaneous luminosity. For this reason, a well-controlled ramp up of instantaneous luminosity may be necessary.
- **Detector performance:** With increased instantaneous luminosity and event pileup, the performance of some sub-detectors degrades due to the effects of out-of-time pileup ('spill-over'), which effectively introduces uncorrelated detector hits in the subsequent bunch crossings and therefore increases the effect of combinatorics and degrades the resolution. Some sub-detectors which are designed for limited channel occupancy may also have intrinsic dead-times at the level of individual channels. At excessive event pileup, this effect acts in the opposite sense of spill-over by introducing detector inefficiencies in the subsequent bunch crossings.
- **Event size:** The event size is proportional to the pileup of physics events together with the hits from background and the spill-over from previous crossings. The readout system may have buffer size limitations which either results in event truncation or dead-time.
- **Trigger rate:** Ideally the trigger rate should just be a product of the physics cross-sections of interest and the instantaneous luminosity. However, at high pileup the trigger selections are strongly affected by the effect of wrongly combining hits or clusters from different events or from spill-over. With increased event pileup, the fake rate at the first level trigger increases rapidly, and the aggravated combinatorics increase very rapidly the average CPU time in the software-based high-level triggers. Some hadronic trigger rates grow exponentially with event pileup (Fig. 2). The excessive rates and processing times ultimately lead to data taking dead-time, which may scale non-linearly with the increased luminosity. Typically, the condition has to be controlled by tightening the selection cuts, often on the requirements on the transverse momentum or energy, with the consequence of decreasing significantly the selection efficiency for the physics. Unlike, for instance, bandwidth limitations which normally affect all physics signals by the same amount, this will affect the efficiency in a physics dependent way. Thus, the additional delivered luminosity offsets the effective usable luminosity by an amount which may be partial, complete, or even worse for low-$p_T$ physics. High trigger efficiency for the Higgs physics while keeping the trigger rate within the budget was one of the biggest challenges for ATLAS and CMS in 2012.

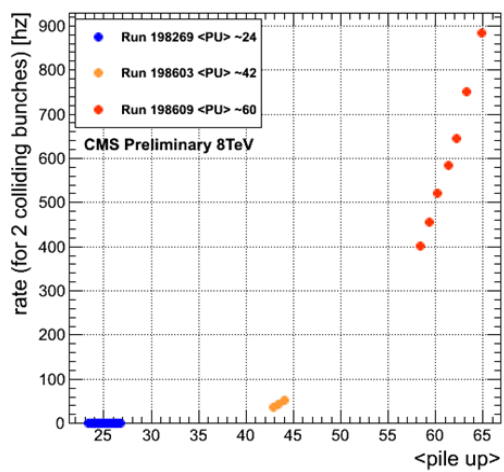

Figure 2: The plot shows the rate per bunch pair of a first level hadron trigger as a function of the pileup. Due to the increasing combinatorics, the fake rate grows exponentially.

- **Readout bandwidth:** The combined effect of increased event size and trigger rates with higher pileup may hit other technical limits related to the readout bandwidth, with similar consequences of introducing dead-time or inefficiencies.
- **Offline processing time:** As in the high-level trigger, the aggravated level of combinatorics with higher pileup increases the reconstruction times almost exponentially. This effect has been mitigated enormously during Run 1 and a lot of work is still in progress to further optimize the code, in particular by profiting efficiently from the parallel architectures of modern CPUs.
- **Machine limitation:** For completeness it should also be mentioned that there could be temporary luminosity limitations introduced by the machine for which it is necessary to operate at a lower instantaneous luminosity than what is potentially available. An example is the current cryogenics limit in IP1 and IP5 with respect to the luminosity debris at an estimated luminosity of ~$1.7 \times 10^{34}$ cm$^{-2}$s$^{-1}$.

- **Luminosity lifetime:** In order to maximize the integrated luminosity, it is interesting to explore the possibility that there may be a potential gain in the luminosity lifetime at the operational limit of the machine by initially levelling the instantaneous luminosity in order to reduce for instance the emittance growth and the collimation burn-off at the peak instantaneous luminosity (Fig. 3). This mechanism should also be combined with the potential increase in physics efficiencies as a consequence of less peak pileup for the trigger, the reconstruction, and the physics resolutions.

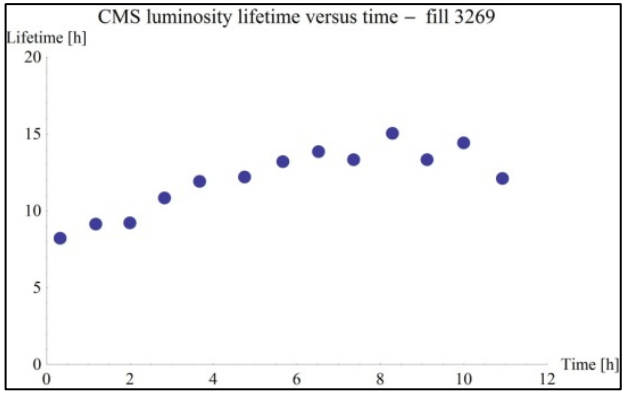

Figure 3: The plot shows an improvement in the luminosity lifetime as the luminosity drops which is partly due to the decreasing rate of collision burn-off. It is interesting to investigate if there are additional effects which may be reduced by levelling the luminosity below its peak value, hence leading to a potential gain in integrated luminosity.

- **Physics:** Even at high pileup, more often than not, the interesting physics event is accompanied by mostly a large number of minimum bias events. Nevertheless, at the limit of the performance, the additional particles from the other primary vertices have the effect of degrading the physics resolutions (vertexing, b-tagging, tracking, momentum, energy [particle-, jet-, missing-], particle identification, etc). The consequences for the physics are reduced background rejection and poorer measurements of the signal. In addition, the systematic errors in precision measurements are often sensitive to changing pileup and temporal variations in the detector performance, trigger, ageing etc. Controlling and levelling the instantaneous luminosity allows working with stable running conditions over months, and makes it easier to monitor and calibrate the detector performance over time.

The last point also elucidates the importance of the size, shape and stability of the luminous region. Each primary vertex is a potential contributor to the usable luminosity. The definition of a usable primary vertex, and the means to immunize the resolutions of the detector observables against pileup, are mainly by associating the hits, energy clusters, tracks and secondary vertices unambiguously to their primary vertices in order to perform pileup suppression and reconstruct each interesting interaction fully. For this reason, a dilution of the primary vertices over the luminous region, typically longitudinally, greatly enhances the situation for the experiment. Obviously the limit in the length of the luminous region at the other extreme is dictated by the finite length of the vertex/inner detectors. A too long luminous region will introduce reconstruction inefficiencies which in turn lead to reduced acceptance. Currently the luminous region is of the order of 50 mm.

Alternatively, the pileup dilution may be enhanced by aiming for a longitudinal flattening of the luminous region away from Gaussian. As a figure of merit for the pileup dilution, the line density as in the average number of events per millimetre is used. However, numerically this number does not reflect well the effect on the usable luminosity with different schemes and shapes. A more appropriate figure of merit should be based on the fraction of the number of primary vertices which may be, or which may not be resolved according to the definition of usable primary vertices above.

In 2012, ATLAS and CMS operated with a peak event pileup of 35 and an average pileup of 20, and LHCb with a constant levelled pileup of 2.1. Significant amount of work went into improving and optimizing the stability of the observables as a function of pileup (Fig. 4).

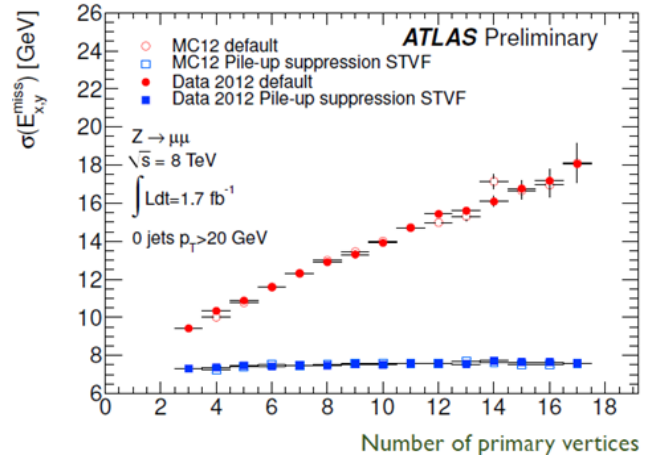

Figure 4: The resolution for missing energy as a function of the pileup. Before applying multivariate techniques to perform pileup suppression, the soft hadrons from the other interactions degraded severely the resolution. Missing energy will play an important role in the characterization of the newly found 126 GeV/$c^2$ neutral scalar boson.

The future priorities will be even more challenging. For ATLAS and CMS, the studies of the Higgs decay channels to $b\bar{b}$ and $\tau\tau$, and the production modes with couplings to W and Z will require even higher pileup immunity in the resolutions of jets, missing energy and $\tau$-lepton identification.

At a given total instantaneous luminosity $L$, the average number of interactions per crossing $\mu$ is given by $\mu =$

$L\sigma/(f_{rev}n_{bb})$, where $\sigma$ is the cross-section, $f_{rev}$ is the LHC revolution frequency and $n_{bb}$ is the number of colliding bunches. The number of interactions per visible crossings, i.e. pileup, is given by $\mu/(1 - e^{-\mu})$. From this it follows that the difference between running with the maximum LHC filling scheme at 50 ns and 25 ns is a factor of two in the event pileup. Of course, 25 ns operation increases the spill-over effect, but this is overall negligible compared to a factor of two in the event pileup. For all the reasons given above, it is then clear that the only path to maximum physics yield is operating at 25 ns. The only scenario which could temporarily outweigh this strategy is either a strong limitation on the instantaneous luminosity at 25 ns from a total intensity limit, electron cloud, etc., or a severely degraded machine stability, i.e. reduced machine time in collision, requiring some longer-term upgrades. For this case, the 50 ns option is considered a backup and a trigger tuning will be prepared. However, maximizing the physics yield at 50 ns will entail considerable changes in the trigger and reconstruction strategies.

While the 50 ns backup option requires levelling for all the IPs with certainty, 25 ns operation should initially only require levelling for ALICE and LHCb. However, ATLAS and CMS are likely to require levelling, if not before, at the latest in the run with High Luminosity LHC (HL-LHC) [3]. At HL-LHC it is expected that the levelling for ATLAS and CMS should allow reducing the potential luminosity of $2 \times 10^{35}$ cm$^{-2}$ s$^{-1}$ by a factor of four.

The next section explores the different options for controlling the luminosity.

## LUMINOSITY/PILEUP CONTROL

Writing the luminosity as

$$L = \frac{n_{bb} \times N^2 \times f_{rev}}{A} * R(\beta^*, \theta, \sigma_z, \phi_P, \delta_S, \delta_C, \Delta t)$$

where $n_{bb}$ is the number of colliding bunches, $N$ is the number of protons per bunch, $f_{rev}$ is the LHC revolution frequency and $A$ represents the physical head-on beam overlap area. $R$ is an overall luminosity factor from optical and geometrical effects which shows that there are many ways to control the luminosity ('luminosity' is here used to denote both instantaneous luminosity and pileup) at an interaction point. While the luminosity control in the ALICE and in the LHCb experiments was based on adjusting the transversal overlap of the two beams in the plane orthogonal to the crossing plane, i.e. with a beam offset in the separation plane ($\delta_S$), there are a number of reasons to consider the other options for the case where all the experiments require luminosity control. An important reason comes from beam stability with lack of Landau damping from head-on collisions [4].

From the experiments' most exigent point of view, the method by which the luminosity is controlled should be local and individual to the experiment and should have minimal impact on the other experiments. Ideally, even ATLAS and CMS could benefit from a decoupled luminosity. Due to the importance of the size and stability of the luminous region, the method by which the luminosity is controlled should not change its length and position significantly. Ideally it should rather maximize the length. The method should involve a minimum of operational overhead and should not limit the flexibility for the other operational modes. The luminosity control should be performed in a way which is safe to the experiment such that it can be done with the experiments fully switched on in the restricted beam mode which is reserved for the safe physics data taking ('Stable Beams'). Any other way would involve 5–10 min loss of physics data taking for each luminosity re-optimization to put the experiments in a safe state and back operational again. The luminosity control should be stable and relatively fine adjustable to allow the luminosity to be maintained within 5% of the desired instantaneous luminosity. This is particularly important for LHCb, but is somewhat less critical to ALICE, ATLAS and CMS. At the optimal limit of the readout and the trigger capacity, luminosity variations either introduce dead-time if the luminosity exceeds the optimal value, or leads to non-optimal running if it is below. Furthermore, since the optimal luminosity may depend on the data taking configuration and the data taking conditions, and even temporary technical limitations, the luminosity should be remotely controllable. This has been of particular importance to the LHCb experiment. Last but not least, the levelling parameters and the related quantities should be measurable and monitored to allow fast analysis of undesired effects.

With the requirements above in mind, below is a list of methods to control the luminosity together with the main positive and negative implications for the experiments.

- **Bunch crossing frequency:** This is not a method to control the luminosity but since it has been suggested as a means of limiting the event pileup at higher luminosity in the future, the option is included here for completeness. The idea aims at distributing the luminosity on more bunches by reducing the bunch spacing and increasing the total number of bunches $n_b$ in the LHC filling scheme. At the same luminosity, it reduces the pileup by $n_b(>40 \text{ MHz})/n_b(40 \text{ MHz})$, but it also increases the stored energy in the LHC by $\sqrt{n_b(>40 \text{ MHz})/n_b(40 \text{ MHz})}$ as compared to the current configuration. Furthermore, it should be understood that the entire readout systems of the experiments are entirely based on signal processing at the level of 25 ns. It would require a complete remake of the whole electronics system to benefit from a higher bunch crossing rate. Moreover, out-of-time pileup effects will be even more important making the gain unclear. It is not considered a viable option.
- **RF cogging ($\Delta t$):** Shifting the RF phase of the beams together with a crossing-angle at the IPs leads to an out-of-time encounter which effectively reduces the luminosity but which also shifts the longitudinal centre of the luminous region by a fraction of the time difference depending on the beam emittance. However, this is not considered a viable option since it affects the luminosity at all the IPs.

- **Crossing angle ($\theta$):** Changing the crossing-angle allows a very limited lever arm within the acceptable range of angles for controlling the luminosity. In particular at small $\beta^*$, the minimum crossing angle is large in order to have sufficient beam separation against long-range interactions. In addition, it is expected that the scheme introduces operational complications with respect to the collimator and orbit management. It is not considered a viable option since it makes the luminous region very short and varying in length during the fill.

- **Bunch rotation with crab cavities ($\phi$):** Equivalent to a crossing-angle, the luminosity may be controlled by rotating the bunch [5]. While the crab cavities are necessary to maximize the luminosity at small $\beta^*$ and maximize the length of the luminous region, their use in levelling is not considered viable since it again reduces strongly the length of the luminous region. For instance, in the nominal 25 ns HL-LHC situation [6], a rotation of 700 µrad would be necessary to reduce the luminosity by a factor four to $5 \times 10^{34}$ cm$^{-2}$ s$^{-1}$, resulting in a luminous region length of <20 mm.

- **Piwinski angle $\phi_P$ (angle $\theta$ + bunch length $\sigma_z$):** The luminosity may also be controlled by changing the Piwinski angle, which involves both the crossing (or crab rotation) angle and the bunch length according to $L \propto 1/\sqrt{1 + [\theta \sigma_z / 2 \sigma_{x,y}]^2}$. This allows a partial compensation against the length shortening of the luminous region by increasing the bunch length. However, this is not considered a viable option since the lever arm to control the luminosity is very limited together with an insufficient compensation for the shortening of the luminous region. In addition, the increase of the bunch length affects all IPs.

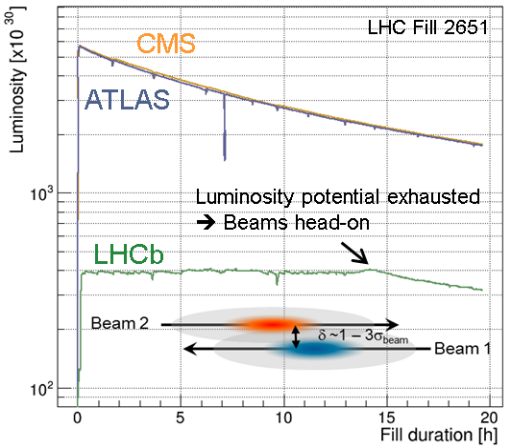

Figure 5: Example of luminosity levelling by separation in LHCb which shows the effect of small orbit drifts and orbit jumps.

- **Beam separation orthogonal to the crossing plane ($\delta_S$):** The luminosity control by adjusting the transversal beam overlap has been successfully used in LHCb and ALICE in Run 1. The operational scheme is simple and the luminous region remains stable both longitudinally and transversally. A drawback is the high sensitivity to orbit drifts and orbit jumps which lead to luminosity instabilities. There is also the risk of accidentally delivering very high instantaneous luminosity head-on. Figure 5 shows an example of the LHCb luminosity in a long fill. While it has been shown that the beam stability tolerates separation of up to a beam sigma [4] even in ATLAS and CMS, and the LHCb and ALICE bunches with larger separation have been stabilized by the head-on Landau damping when colliding in ATLAS and CMS, the main concern is the complete lack of Landau damping if this scheme is applied in all experiments. The required beam separations are not expected to introduce instabilities but the lack of head-on collisions means that there is no Landau damping of instabilities from other sources. In general the scheme is considered a viable option by the experiments.

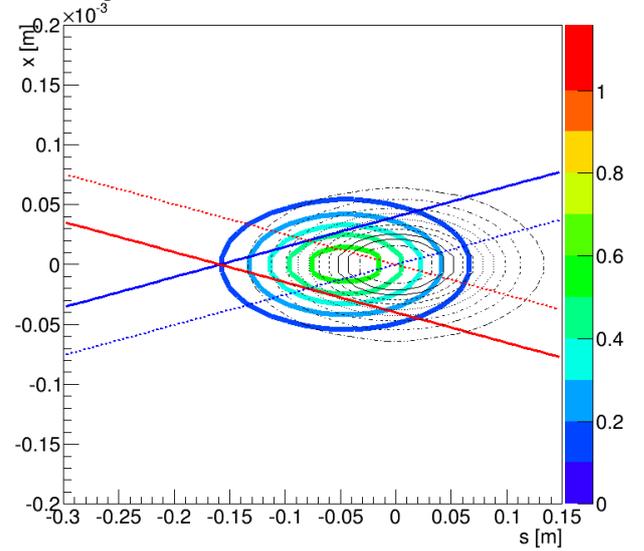

Figure 6: Example of luminosity levelling by separation in the crossing plane. The dashed lines show the luminous region and the geometrical crossing point with head-on collision while the solid lines show the luminous region and the geometrical crossing point in LHCb with the Run 2 beam parameters and a shift of one sigma shift per beam. The shift leads to a luminosity reduction of a factor two and shifts the centre of the luminous region by almost 5 cm.

- **Beam separation in the crossing-plane ($\delta_c$):** Shifting the beams laterally in the crossing-plane also allows controlling the luminosity with a limited length shortening of the luminous region and a limited shift of the longitudinal centre of the luminous region which depends on the angle and the beam emittance. If the beam emittance is very small, the centre is close to the geometrical crossing point while if the beam emittance is large the centre remains near the beam encounter at $t = 0$. Figure 6 shows an example with the configuration for LHCb in Run 2 and a shift of both beams by a beam sigma resulting in a luminosity reduction of factor of two. Again this option is not preferred since it shifts the centre of the luminous region. Nevertheless it would be interesting to check if the two types of separation have the same effect on the beam stability.

- **Beam focusing ($\beta^*$):** Controlling the luminosity by the $\beta^*$ is clearly a more involved scheme which requires new levels of control and a scheme to manage the optics and the collimators without leaving the 'Stable Beams' condition [7]. However it has the advantage that it imposes no limitation on the luminosity range, it preserves the luminous region longitudinally, and it is relatively immune to orbit variations. The $\beta^*$ squeeze in Stable Beams to compensate for the emittance growth and the intensity drop could also be performed in only one plane while maintaining the $\beta^*$ small in the other. This would have the interesting feature of preserving the luminous region transversally in one plane, and supports the idea of the use of flat beams in the future [8]. If the $\beta^*$ squeeze in Stable Beams cannot be smooth, the scheme could be combined with one of the beam separation schemes above to level the luminosity between the successive $\beta^*$ steps. The combined scheme could be constructed to maintain the range of separation below one beam sigma to respect the stability diagram. There is some concern about the limit from the total head-on beam–beam tune shift if this scheme is applied in three IPs. However, as of now no limit has been observed and the limit has been demonstrated to be well beyond the design expectations [9].

Table 2 shows a list of examples of luminosity levelling by purely transversal separation and by beam focusing in both planes in the future runs in order to give an indication of the required ranges. The minimum $\beta^*$ in LHCb has been chosen in order to guarantee a levelling lifetime of >12 h assuming similar emittance and intensity lifetimes to what was achieved in Run 1. The head-on luminosity in ALICE is of concern with separation as it could potentially induce damage. The main background monitor which is connected to the LHC beam dump has thresholds adjusted to a level which dumps the beam in case of excessive luminosity. Furthermore, with levelling by separation in ATLAS and CMS in Run 2, orbit variations of $\pm 1$ μm at the IP change the luminosity and pileup by as much as 8%.

All of these arguments appear to favour a combined levelling by beam focusing and a restricted transversal separation. If luminosity levelling by beam focusing is performed in only one plane, the maximum $\beta^*$ required in the levelling plane is trivially $\beta_l^* = \beta_{max}^{*2}/\beta_{min}^*$ as compared to the numbers in Table 2. Clearly, the addition of a combined transversal separation allows reducing the maximum $\beta^*$ required.

| | | ATLAS/CMS | LHCb | ALICE |
|---|---|---|---|---|
| Run 2 | Mimimum β* | 0.4 m | 3 m | 10 m |
| | Head-on luminosity [cm$^{-2}$s$^{-1}$] | 1.5×10$^{34}$ | 3×10$^{33}$ | 1×10$^{33}$ |
| | Levelled luminosity [cm$^{-2}$s$^{-1}$] | No | 4-6×10$^{32}$ | 4×10$^{30}$ |
| | Level by separation δ$_s$ at β*$_{min}$ | – | 2.8σ → 0σ | ~5σ |
| | Level by β* at δ$_s$=0 | – | 20m → 3m | – |
| Run 3 | Mimimum β* | 0.4 m | 3 m | 10 m |
| | Head-on luminosity [cm$^{-2}$s$^{-1}$] | 2.5×10$^{34}$ (p.u.~70) | 5×10$^{33}$ | 1.5×10$^{33}$ |
| | Levelled luminosity [cm$^{-2}$s$^{-1}$] | 1.5×10$^{34}$ (example) | 1-2×10$^{33}$ | 2×10$^{31}$ |
| | Level by separation δ$_s$ at β*$_{min}$ | 1.5σ → 0σ | 2.5σ → 0σ | ~4σ |
| | Level by β* at δ$_s$=0 | 0.8 m → 0.4 m | 15 m → 5 m | – |
| HL-LHC | β* | 0.15 m + crab cavity | 3 m | 10 m |
| | Head-on luminosity [cm$^{-2}$s$^{-1}$] | 2.4×10$^{35}$ | 1×10$^{34}$ | 3.5×10$^{33}$ |
| | Levelled luminosity [cm$^{-2}$s$^{-1}$] | 5×10$^{34}$ (pu~140) | 2×10$^{33}$ | 2×10$^{31}$ |
| | Level by separation δ$_s$ at β*$_{min}$ | 2.5σ → 0σ | 2.5σ → 0σ | ~4.5σ |
| | Level by β* at δ$_s$=0 | 0.7 m → 0.15 m | 16 m → 3 m | – |

Table 2: Examples of luminosity levelling by purely transversal separation and by beam focusing in both planes in the future runs assuming the preferred option of 25 ns operation with the parameters in Table 1.

# SUMMARY OF EXPERIMENT DESIDERATA FOR THE FUTURE

As a starting point on the running conditions for the future it should be generally noted that already at the same instantaneous luminosity, the event complexity increases naturally when increasing the energy from 8 TeV to 13 TeV due to two effects. The minimum bias cross-section and the associated event pileup is expected to increase by about 15% and the overall multiplicity is expected to increase by ~10–20% depending on the rapidity range. Thus at the same luminosity and the same number of colliding bunches, the events get ~25% more busy in ATLAS and CMS and up to ~40% more busy in LHCb.

For 2015, the experiments assume an initial period of 50 ns operation in order to restart the LHC, commission 13 TeV physics, and for the initial electron cloud scrubbing and the intensity ramp up to a full 50ns machine. The experiments will profit from this phase to perform calibrations and re-commission the detectors and the entire data flow chain. Of particular importance is the commissioning and validation of the triggers and the offline processing which are being extensively revised during LS1 in order to cope with higher pileup. It is expected that the first goal will be to commission the triggers tuned for 25 ns, which in itself may mean limiting the pileup for a short period to the maximum value expected with 25 ns operation. This is obviously only necessary if LHC is potentially already able to deliver significantly higher pileup at 50 ns. It should still be noted that this phase does not fully allow assessing the effect of spill-over in the trigger and reconstruction which introduces some uncertainty in the final tuning. However, this is addressed in parallel by simulation.

Secondly, it is expected that the experiments will be commissioning a backup trigger tuning for 50 ns operation which can then also be used to collect as much integrated luminosity as possible in this first phase.

Nevertheless, the experiments would like to see this phase as short as possible, also implying that no extra time should be spent on optimizing the luminosity in this phase beyond what is necessary to allow the transition to 25 ns operation.

The transition to 25 ns is vital for all the reasons given above and should be the primary goal. It is understood and accepted that this may require a significant amount of commissioning and conditioning time in 2015. It is also clear that the higher stored energy may lead to machine instabilities and may also contribute to limiting the initial instantaneous luminosity at 25 ns.

Clearly the progress will have to be followed closely and some breakpoints for decisions on the strategy will be needed, but in general this phase is considered an investment for the future.

A special case is if the LHC injectors are not able to deliver the 25 ns-BCMS scheme efficiently to fill the entire LHC machine. An instantaneous luminosity limit of <$10^{34}$ cm$^{-2}$ s$^{-1}$ at 25 ns with no prospect for improvement may be considered a tipping point for ATLAS and CMS. Since ALICE and LHCb will be running with a levelled instantaneous luminosity, a tipping point is *only* valid if the machine availability is severely affected at 25 ns. This particular case will therefore require a difficult assessment with all elements at hand on the moment.

In the preferred scenario with 25 ns operation, ATLAS and CMS are not a priori expected to require levelling of the luminosity in Run 2, in particular with the upcoming upgrades. Currently this assumes that the luminosity remains below $2 \times 10^{34}$ cm$^{-2}$ s$^{-1}$. This means effectively that ATLAS and CMS expect to be able to cope with a peak pileup of up to 45–60 in Run 2 and Run 3. In this configuration, ATLAS and CMS should be able to collect realistically about 100 fb$^{-1}$ in Run 2 and another 300 fb$^{-1}$ in Run 3 with the upgraded LHC injectors.

In case 50 ns operation turns out to be significantly more productive than 25 ns, luminosity levelling will be needed even with the nominal 50 ns beam parameters. The exact level of pileup at which ATLAS and CMS require levelling of the luminosity is currently under study.

LHCb will require levelling of the instantaneous luminosity from the first day of luminosity production in 2015. In the preferred scenario with 25 ns operation, the LHCb target luminosity is expected to be around $4 \times 10^{32}$ cm$^{-2}$ s$^{-1}$ for Run 2. After the LHCb upgrade in LS2, LHCb is expecting to be initially running at a levelled luminosity of $1 \times 10^{33}$ cm$^{-2}$ s$^{-1}$ in Run 3, and later move to a levelled luminosity of $2 \times 10^{33}$ cm$^{-2}$ s$^{-1}$. In all cases, the $\beta^*$ configuration should be such that it allows LHCb to run at constant luminosity with a levelling lifetime which is of the order of the longest typical fill duration (10–15 h). In these conditions, LHCb should be able to collect realistically about 5 fb$^{-1}$ in Run 2 and aim at 50–100 fb$^{-1}$ after the upgrade in LS2. Like with ATLAS and CMS, the studies of the exact conditions for the backup case of 50ns operation has not been completed.

While luminosity levelling is a priori not needed by ATLAS and CMS in the preferred scenario for Run 2, it is strongly felt that a levelling scheme should be prepared. On the one hand, it could be used in case of the unforeseen; on the other hand it allows studying the feasibility for Run 3 and for HL-LHC. For this reason, the choice of levelling method should be the most promising for HL-LHC.

For the reasons outlined in the previous Sections, luminosity levelling by $\beta^*$ and by transversal separation are considered the two most viable options by the experiments. The levelling by separation is to a large extent already available. It is felt that the levelling by $\beta^*$ should be studied and implemented, and that the most appropriate option is likely to be a combined levelling by $\beta^*$ with transversal separation to smoothen the luminosity between the successive $\beta^*$ steps within a limited range which still provides Landau damping.

Consequently, there is strong interest in putting $\beta^*$ levelling in operation for LHCb for Run 2. Head-on collision in LHCb should also allow exploring further the beam–beam limit for the future.

At HL-LHC, the optimal configuration for ATLAS and CMS appears to be a combined $\beta^*$/separation levelling scheme with crab cavities providing constantly a bunch rotation that gives a complete bunch-bunch overlap in the crossing plane to maximize the length of the luminous region.

ALICE will need to take proton–proton data yearly for physics normalizations and detector calibrations, but also for operational preparation and verification for the ion runs. In Run 2 with 25ns, ALICE plans to run in two different trigger configurations to collect a minimum of about 3 pb$^{-1}$/year of barrel triggers at a levelled luminosity of $4 \times 10^{30}$ cm$^{-2}$ s$^{-1}$ and about 15 pb$^{-1}$/year of muon triggers at a levelled luminosity of $2 \times 10^{31}$ cm$^{-2}$ s$^{-1}$. In the backup case of 50 ns operation, ALICE would request about 45 isolated main–main collisions per turn with a levelled luminosity as the ideal data taking condition. In Run 3 after the upgrade in LS2, ALICE expects to require a minimum of 100 pb$^{-1}$ of proton–proton data per year at a levelled luminosity of $2 \times 10^{31}$ cm$^{-2}$ s$^{-1}$. It is expected that the levelling in ALICE will continue to be performed by transversal beam separation. The target luminosity for Run 2 and 25 ns operation implies a separation of about 5 $\sigma$ at a $\beta^*$ of 10 m. However, with such a large separation, accidental excessive luminosities are a concern since the beams will be dumped and there is a small risk of detector damage. It seems interesting to investigate the possibility of running at a larger $\beta^*$ to reduce the maximum potential luminosity.

ATLAS, CMS, LHCb and TOTEM all depend critically on the length of the luminous region to resolve the primary vertices, and on the longitudinal stability. In general the luminous regions should be maintained long and an increase of 10–15% could improve the situation for all experiments. Since the size is a parameter in the physics simulations, the value should be known well in advance and then kept as stable as possible. In addition, in order to improve the acceptance for long-lived B hadrons, LHCb could benefit from a small upstream (anticlockwise) shift of the centre of the luminous region. This may be achieved by a simple small lateral shift of the two beams in the crossing-plane as shown in Fig. 6. Studies of the optimal configuration are underway.

Non-colliding bunches were used by ATLAS, CMS and LHCb in Run 1 to measure single-bunch related physics background and for background subtraction in the luminosity determination. In the nominal 25 ns filling scheme there is no longer room for non-colliding bunches in ATLAS and CMS. Nevertheless, assuming that a stable scheme with negligible impact on luminosity can be found, ATLAS and CMS would still like to have a few non-colliding bunches. LHCb will still get non-colliding naturally at the end of the trains due to the 11.25 m shift from the position of the nominal IP.

LHCb needs to continue operating with the combination of a horizontal internal crossing-angle from the spectrometer magnet and an external vertical angle in order for the effective net crossing-angle to be the same in both polarities. At 25 ns the negative polarity of the spectrometer magnet results in parasitic long-range collisions without the vertical external angle. This means that the tilted crossing scheme has to be set already at injection. This is currently part of a detailed study [10].

As in Run 1, ALICE and LHCb will also require the regular polarity reversals.

In addition to the special high-$\beta^*$ runs, TOTEM will take data together with CMS during the high-luminosity proton–proton runs in Run 2 and in Run 3. It is hoped that TOTEM will be able to move the horizontal pots to 14$\sigma$ and the vertical pots to 11 $\sigma$ in the high-luminosity runs in Run 2. Luminosity levelling by $\beta^*$ adds a complication to TOTEM in that corrections have to be applied in the analyses for the changes of the optics.

LHCf plans to take data together with ATLAS in Run 2 during the very first phase of proton–proton physics. It will be for a very limited amount of time as the detector ageing is expected to degrade the performance after about 1 fb$^{-1}$. For this reason, the special runs for LHCf should be scheduled very early in 2015, ideally before 500 nb$^{-1}$ has been accumulated which effectively means during the machine commissioning phase. The requested conditions for the data taking consist of less than 40 colliding bunches and injection optics not to exceed a luminosity of $10^{29}$ cm$^{-2}$ s$^{-1}$, and an integrated luminosity of 10 nb$^{-1}$ at several centre-of-mass energies.

## CONCLUSION

LHC proved to be an extremely versatile machine in Run 1. While the experiments challenged the machine with a very wide range of requirements, the machine challenged the experiments with unrefusable running conditions which at the end resulted both in physics results of fundamental importance, but also operational prospects of fundamental importance for the future runs.

This paper recapitulates on this outcome, and motivates and outlines the future constraints, requirements and preferences from the experiments which need to be taken into account when defining the future strategy. Many of these requirements are challenging from the point of view of beam–beam effects and will require careful assessment. This is clearly an iterative process between the experiments and the machine which will eventually define a baseline for the running configuration and conditions which maximize the physics performance. Many studies are currently in progress in the experiments to refine the requirements and the limits for the future. It is expected that these studies will be able to deliver more accurate statements at the beginning of 2014.

Nevertheless, it's more than likely that there will be as many surprises and new ideas flowing in the future as in Run 1, which ultimately means that both the experiments and the machine will have to stay flexible.

## ACKNOWLEDGMENTS

This paper cannot be concluded without thanking all the colleagues from the LHC and injectors for the exceptional performance of the machines and the fantastic collaboration.

This paper has been presented on the behalf of ALICE, ATLAS, CMS, LHCb, LHCf and TOTEM.

The author would like to thank the organizers of the conference for the invitation to give this talk and thank in particular G. Arduini, O. Bruning, T. Camporesi, M. Chamizo, M. Deile, A. Di Mauro, B. Gorini, W. Herr, M. Lamont, E. Meschi, G. Papotti, T. Pauly, T. Pieloni, G. Rakness, T. Sako and S. Zimmermann for providing input and comments to this paper.

## REFERENCES

[1] S.S. Gilardoni, "The High Intensity/Intensity Brightness Upgrade Programme at CERN: Status and Challenges," Proc. 52nd ICFA Advanced Beam Dynamics Workshop on High-Intensity and High-Brightness Hadron Beams, Beijing, 2012.

[2] R. Steerenberg et al., "Post-LS1 25 ns & 50 ns Options from the Injectors," Proc. LHC Beam Operation Workshop, Evian, France, 2012.

[3] HL-LHC, http://hilumilhc.web.cern.ch/HiLumiLHC/

[4] X. Buffat, "Coherent Beam–Beam Modes in the LHC," these proceedings.

[5] G. Sterbini, "Luminosity Levelling with Crabs," Proc. 4th LHC Crab Cavity Workshop, CERN, Geneva, 2010.

[6] O.S. Bruning and F. Zimmermann, "Parameter Space for the LHC Luminosity Upgrade," Proc. IPAC2012, New Orleans, Louisiana, USA.

[7] X. Buffat et al., "Results of $\beta^*$ luminosity levelling," (CERN ATS-Note-2012-071 MD, 2012).

[8] W. Herr et al., "Performance Reach of LHC after LS1," Proc. LHC Performance Workshop, Chamonix, France, 2012.

[9] T. Pieloni et al., "Colliding High Brightness Beams in the LHC," Proc. 52nd ICFA Advanced Beam Dynamics Workshop on High-Intensity and High-Brightness Hadron Beams, Beijing, 2012.

[10] B.J. Holzer, "Spectrometer Operation in IR2 & IR8," Proc. LHC Beam Operation Workshop, Evian, France, 2012.